\documentclass[letterpaper]{article} 
\usepackage{aaai2027}  
\usepackage[hyphens]{url}  
\usepackage{graphicx} 
\usepackage{natbib}
\usepackage{caption} 
\usepackage{amsmath,amssymb,amsthm}
\usepackage{mathtools}
\usepackage{booktabs}
\usepackage{multirow}
\usepackage{algorithm}
\usepackage{algorithmic}
\usepackage{array}
\usepackage{placeins}

\newtheorem{theorem}{Theorem}

\newtheorem{definition}{Definition}

\newcommand{\method}{\textnormal{\textsc{FedWorld}}}
\newcommand{\localwm}{\textnormal{\textsc{LocalWM}}}
\newcommand{\naiveunion}{\textnormal{\textsc{NaiveUnion}}}

\title{FedWorld: Scope-Aware Federation of Agent World Models}

\author{Yuchao Hou}
\affiliations{Shanxi Key Laboratory of Cryptography and Data Security, Shanxi Normal University, Taiyuan, China}
\begin{document}
	\maketitle
\begin{abstract}
	Large language model (LLM) agents learn world dynamics from local interaction experience to support subsequent planning and action selection. However, the experience available to a single client is often incomplete, which motivates sharing knowledge across clients. Existing federated methods mainly aggregate model parameters, while agent memory-sharing methods commonly pool trajectories, memories, or rules without checking whether they remain valid for each client. This assumption is problematic because the same abstract action may produce different effects under different policies, environments, or exception conditions. Consequently, a rule supported by most clients may overwrite correct knowledge held by a minority client. To address this problem, we propose \method{}, a scope-aware federated world-model protocol that exchanges structured abstract transition rules. Each client converts private transitions into normalized rules, and the server aligns related rules to identify each rule supporting and contradicting evidence across clients. The resulting evidence determines whether a rule is shared, cluster-specific, private, or unresolved. Each target client retains its local rules and accepts federated updates only for uncovered cases whose inferred scope is compatible; ambiguous rules are withheld. Experiments on $\tau$-bench and ALFWorld show that \method{} reduces negative transfer under conflicting dynamics while retaining useful cross-client transfer, leading to fewer state regressions, repeated actions, and excess steps, as well as higher task success.
\end{abstract}
	
	\section{Introduction}
	\label{sec:intro}
	
Large language model (LLM) agents increasingly use past interactions to predict what their actions will change. A retail-support agent, for example, must know whether cancelling an order after shipment leads to a refund or store credit, while an embodied agent must know how moving or heating an object changes its state. The same need arises in web navigation and software tasks, where one incorrect state prediction can mislead all later actions~\citep{liu2024agentbench,qin2024toolllm,zhou2024webarena,deng2024mind2web,jimenez2024swebench,yao2025taubench,shridhar2021alfworld}. We use \emph{world model} to describe this persistent knowledge of how a state and an action lead to an effect. This meaning follows model-based reinforcement learning, where transition knowledge supports planning and policy improvement~\citep{ha2018worldmodels,hafner2019planet,hafner2020dreamer,schrittwieser2020muzero}. LLM agents may store such knowledge in trajectories, reflections, memories, retrieved records, or symbolic action rules~\citep{yao2023react,shinn2023reflexion,zhao2024expel,park2023generativeagents,lewis2020rag}.

However, the transition knowledge stored by a single client is usually incomplete. Each client encounters only a limited range of policies, exceptions, object states, and action outcomes, so other clients may hold rules that fill its missing cases. Federation offers a way to share this knowledge without centralizing raw trajectories. Existing federated learning methods mainly achieve such collaboration by aggregating model parameters~\citep{mcmahan2017fedavg,li2020fedprox,karimireddy2020scaffold}, while personalized methods retain local or clustered components to accommodate client differences~\citep{smith2017mocha,dinh2020pfedme,ghosh2020clusteredfl,li2021ditto}. These methods can combine what clients have learned, but they do not decide whether a specific transition rule from one client also holds for another.
This missing validity check matters because the same action can follow different dynamics across clients. Under one retail policy, cancelling an order may lead to a refund, while another policy may issue store credit. In a household environment, heating, cooling, or moving an object may also produce different effects depending on its current state or receptacle. These cases share the same abstract action but not the same transition rule. If they are pooled without preserving the conditions that separate them, a rule supported by most clients may become incorrect for a minority client.

Cross-client rule sharing therefore has two different outcomes. A compatible rule can fill a missing part of the target client's world model. An incompatible rule can instead replace knowledge that is already correct. Existing agent memory methods mainly retrieve experience according to semantic relevance or similarity~\citep{yao2023react,shinn2023reflexion,zhao2024expel,lewis2020rag}. Such relevance helps identify related situations, but it does not show whether they obey the same transition dynamics. Safe federation must therefore determine both which rules are related and where each rule remains valid. We formulate scope-aware federation of agent world
	models as a distinct federation setting in which the exchanged
objects are discrete transition rules rather than continuous
model parameters. At the rule level, averaging does not resolve
heterogeneity: a transition is either applicable to a target
client under its conditions or it is not. The central objective
is therefore to identify whether each rule is shared,
cluster-specific, private, or unresolved before it can affect
the target client. This formulation separates useful
cross-client gap filling from the negative transfer caused by
scope-incompatible rules.
	
To address this problem, we propose \method{}, a protocol that exchanges structured abstract transition rules rather than model parameters, raw trajectories, or free-form memories. Each client maps its private transitions into normalized rules, which makes related experience comparable across clients. The server aligns these rules and gathers supporting and contradicting evidence from eligible clients. Based on this evidence, \method{} identifies each rule as shared, cluster-specific, private, or unresolved. Each target client uses the inferred scope to admit only compatible federated rules for uncovered cases while preserving its existing local knowledge.	 This allows cross-client experience to fill local gaps without assuming that every shared rule is valid for every client.
	
The main contributions are as follows:
\begin{itemize}
	\item We formulate scope-aware federation of agent world
	models, where clients exchange discrete transition rules
	whose validity may be shared, cluster specific, private, or
	unresolved. This setting makes rule applicability, rather
	than parameter averaging, the central federation objective.
	
	\item We characterize rule-scope mismatch and its resulting
	strict conflict negative transfer, in which a majority supported
	rule can overwrite correct knowledge held by a minority client.
	
	\item We instantiate this formulation with \method{}, which
	aligns abstract transition rules, infers their valid scope, and
	uses a local-first compiler to fill compatible knowledge gaps
	without replacing existing local rules.
\end{itemize}
\section{Related Work}
\label{sec:related}

Federated learning enables multiple clients to benefit from decentralized private data without sharing raw samples. FedAvg establishes the basic parameter-aggregation framework, while later methods address statistical and system heterogeneity through proximal regularization, normalized aggregation, control variates, dynamic regularization, and local feature statistics~\citep{mcmahan2017fedavg,li2020fedprox,wang2020fednova,karimireddy2020scaffold,acar2021feddyn,li2021fedbn}. Personalized federated learning further allows clients to retain clustered, partially shared, or client-specific parameters when a single global model is insufficient~\citep{smith2017mocha,dinh2020pfedme,ghosh2020clusteredfl,li2021ditto}. This decentralized setting also appears when deployed agents build world models from private local interactions. Since each client observes only part of the available policies, exceptions, and state transitions, federating local world models can extend their coverage without centralizing raw trajectories. The shared object, however, is no longer limited to trainable parameters. World models encode how actions change the environment and support planning through predicted state transitions~\citep{ha2018worldmodels,hafner2019planet,hafner2020dreamer,schrittwieser2020muzero}. Language-agent systems represent similar knowledge through reasoning traces, reflections, memories, experience libraries, and retrieved records~\citep{yao2023react,shinn2023reflexion,zhao2024expel,park2023generativeagents,lewis2020rag,karpukhin2020dpr}, while agent benchmarks evaluate its effect on tool use, web navigation, software engineering, and embodied control~\citep{liu2024agentbench,qin2024toolllm,zhou2024webarena,deng2024mind2web,jimenez2024swebench,mialon2024gaia,yao2025taubench,shridhar2021alfworld}. Existing memory and retrieval methods usually select experience by similarity or relevance, but related records may still follow different dynamics across clients. Conventional federated methods likewise do not determine where an explicit transition rule remains valid. In contrast, \method{} exchanges structured abstract transition rules and treats heterogeneity as a scope-identification problem, so a rule affects a target client only when its inferred scope permits transfer.
	
\section{Problem Setup and Scope Formulation}
\label{sec:prelim}

We consider \(K\) clients indexed by
\(\mathcal{K}=\{1,\ldots,K\}\). Client \(k\) observes a private set of transitions,
\begin{equation}
	D_k=\{(s_t,a_t,s_{t+1})\},
	\label{eq:client-trajectories}
\end{equation}
generated under its own environment, policy, or business-rule configuration. Since these transitions may contain client-specific states and arguments, they cannot be compared or exchanged directly. We therefore introduce a normalization map
\begin{equation}
	\phi_k:D_k\rightarrow \mathcal{Z},
	\label{eq:normalization-map}
\end{equation}
which converts each detailed transition into an abstract record
\begin{equation}
	z=(a,x,c,e,\varepsilon)\in
	\mathcal{A}\times\mathcal{X}\times\mathcal{C}\times\mathcal{E}\times\mathcal{R}.
	\label{eq:abstract-record}
\end{equation}
Here, \(a\) denotes the abstract action type, \(x\) the normalized pre-state, \(c\) the normalized condition set, \(e\) the abstract effect, and \(\varepsilon\) the normalized exception structure.

Client \(k\) keeps both the detailed transitions required for local execution and the corresponding abstract rule store:
\begin{align}
	\mathrm{LWM}_k&=(D_k^{\mathrm{detail}},A_k),
	\label{eq:local-world-model}\\
	A_k&=\{\phi_k(d):d\in D_k\}.
	\label{eq:abstract-store}
\end{align}
Only \(A_k\) is shared during federation. Raw trajectories, entity identifiers, tool arguments, detailed states, and the mapping from detailed transitions to abstract rules remain on the client.

Once the local abstract stores are available, we form the federated rule store
\begin{equation}
	F=\bigcup_{k\in\mathcal{K}} A_k.
	\label{eq:federated-store}
\end{equation}
The main difficulty is that membership in \(F\) does not make a rule valid for every client. Let
\(\mathcal{R}_{\mathrm{ok}}=\{\mathrm{match},\mathrm{compatible}\}\).
For an abstract query \(u\) from target client \(k\), naive federation considers all aligned candidates:
\begin{equation}
	F_{\mathrm{all}}(k,u)=
	\{g\in F:\operatorname{rel}(u,g)\in\mathcal{R}_{\mathrm{ok}}\}.
	\label{eq:all-candidates}
\end{equation}
This candidate set captures semantic compatibility, but it does not account for the clients on which each rule is valid. We address this gap by estimating a scope map
\begin{equation}
	\sigma:F\rightarrow
	\{\mathrm{shared},\mathrm{cluster},\mathrm{private},\mathrm{unresolved}\},
	\label{eq:scope-map}
\end{equation}
and retaining only candidates whose inferred scope is compatible with the target client:
\begin{align}
	F_{\mathrm{FW}}(k,u)
	&=\{g:\;g\in F_{\mathrm{all}}(k,u),\notag\\
	&\qquad\operatorname{Compat}(k,\sigma(g))=1\}.
	\label{eq:fedworld-candidates}
\end{align}

\begin{definition}[\method{}]
	\label{def:fedworld}
	We define \method{} as the federated agent world-model protocol
	\begin{equation}
		\method{}\doteq
		(\phi,\operatorname{rel},\sigma,
		\operatorname{Compat},\operatorname{Compile}),
		\label{eq:fedworld-tuple}
	\end{equation}
	where \(\phi\) abstracts private transitions,
	\(\operatorname{rel}\) aligns query and candidate rules,
	\(\sigma\) infers the valid scope of each candidate,
	\(\operatorname{Compat}\) checks its applicability to the target client,
	and \(\operatorname{Compile}\) converts an admissible rule into an executable world-model update.
	A rule in
	\(F_{\mathrm{all}}(k,u)\setminus F_{\mathrm{FW}}(k,u)\)
	may still provide cross-client evidence, but it cannot be used by target client \(k\).
\end{definition}

\section{Proposed Method}
\label{sec:method}

\subsection{Overview and Design Intuition}
\begin{figure*}[!t]
	\centering
	\includegraphics[width=.98\textwidth]{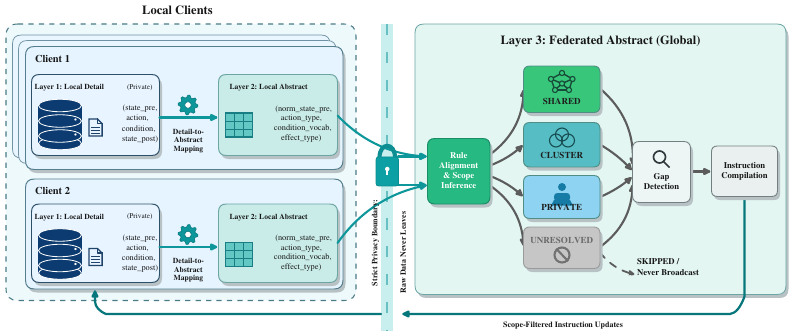}
	\caption{\method{} framework.}
	\label{fig:method_framework}
\end{figure*}

We design \method{} as a local-first, scope-aware protocol for federating agent world models. As shown in Figure~\ref{fig:method_framework}, each client keeps detailed trajectories local and shares only normalized abstract rules. The server aligns related rules and infers their valid scope from cross-client evidence. Since the same action may follow different dynamics across policies or environments, the target client consults federated rules only when its local store lacks coverage. The compiler then admits scope-compatible candidates and withholds unresolved or incompatible rules, allowing cross-client experience to fill local gaps without replacing correct local knowledge.

\subsection{Problem Definition and Objective}

For target client \(k\), \method{} receives the local abstract store \(A_k\), the federated store \(F=\bigcup_j A_j\), and a query rule \(r\). The query specifies an abstract pre-state, action, and condition set, while its effect is unknown. The protocol returns a predicted effect \(\hat e\), a compiled instruction \(\hat I_k(r)\), or abstention.

We write the runtime interface as
\begin{equation}
	x_k \doteq (A_k,F,r,k),\qquad
	\Psi(x_k)\in
	\{\hat e,\hat I_k(r),\mathrm{abstain}\},
	\label{eq:method-io}
\end{equation}
where \(\Psi\) denotes the prediction and compilation procedure. The target index \(k\) is part of the input because the same federated rule may be valid for one client but invalid for another.

The objective balances three requirements. The protocol should predict the correct effect, avoid damaging correct local knowledge, and abstain only when no reliable rule is available:
\begin{align}
	\min_{\Psi}\quad
	&\mathbb{E}_{(k,r,e)}
	\left[\mathbf{1}\{\hat e_{\Psi}(k,r)\ne e\}\right]\notag\\
	&+\lambda_{\mathrm{ntr}}
	\mathbb{E}_{(k,r,e)}
	\left[
	\mathbf{1}\{
	\mathrm{LWM}_k(r)=e,
	\hat e_{\Psi}(k,r)\ne e
	\}
	\right]\notag\\
	&+\lambda_{\mathrm{abs}}
	\mathbb{E}_{(k,r)}
	\left[
	\mathbf{1}\{
	\Psi(k,r)=\mathrm{abstain}
	\}
	\right],
	\label{eq:method-objective}
\end{align}
where \(e\) is the ground-truth abstract effect. The coefficients
\(\lambda_{\mathrm{ntr}},\lambda_{\mathrm{abs}}\geq0\) control the penalties for harmful overwrite and abstention. The harmful-overwrite term is important because a retrieved rule should not replace a local prediction that is already correct.

We impose two constraints to reflect this objective:
\begin{align}
	u\in A_k
	&\Rightarrow
	\hat e_{\Psi}(k,r)=\mathrm{LWM}_k(u),
	\label{eq:local-first-constraint}\\
	g\notin F_{\mathrm{FW}}(k,u)
	&\Rightarrow
	g\notin\operatorname{Compile}_k(u).
	\label{eq:scope-constraint}
\end{align}
Equation~\eqref{eq:local-first-constraint} gives priority to an existing local rule. Equation~\eqref{eq:scope-constraint} prevents a federated candidate from affecting the target unless it passes scope filtering. A rule may therefore remain visible as cross-client evidence without becoming an executable instruction.

Together, these constraints define a target-conditioned world model:
\begin{equation}
	\widehat W_{\method,k}(r)=
	\bigl(
	u,\mathcal{C}_u,\sigma,
	F_{\mathrm{FW}}(k,u),
	\hat e,\hat I_k
	\bigr),
	\label{eq:fedworld-world-model}
\end{equation}
where \(u=\operatorname{key}(r)\), \(\mathcal{C}_u\) is the aligned candidate bucket, \(\sigma\) is the scope function, and \(F_{\mathrm{FW}}(k,u)\) contains the candidates admitted for client \(k\). Thus, \method{} does not simply enlarge the local memory. It controls which cross-client transitions may influence a particular client.

\subsection{M1: Abstract Rule Mapping}

Detailed transitions cannot be compared directly because clients use different entities, arguments, and surface descriptions. We address this mismatch by mapping each transition into a common abstract representation. For rule \(r\), we define the identity key
\begin{equation}
	\operatorname{key}(r)=
	\bigl(
	\operatorname{Norm}(S_{\mathrm{pre}}),
	\operatorname{Norm}(A),
	\operatorname{Norm}(C)
	\bigr),
	\label{eq:method-key}
\end{equation}
where \(S_{\mathrm{pre}}\) is the pre-state, \(A\) is the action, and \(C\) is the condition set. The function \(\operatorname{Norm}(\cdot)\) maps client-specific fields into a controlled abstract vocabulary.

\subsection{M2: Cross-Client Rule Alignment}

The abstract key identifies potentially related rules, but it does not establish that they have the same meaning or effect. We therefore assign a relation label to target key \(u\) and candidate rule \(g\in F\):
\begin{equation}
	\operatorname{rel}(u,g)\in
	\{
	\mathrm{match},
	\mathrm{compatible},
	\mathrm{conflict},
	\mathrm{unresolved}
	\}.
	\label{eq:relation-labels}
\end{equation}

A matched rule describes the same abstract transition, while a compatible rule can be used under the target conditions. A conflict label indicates incompatible conditions or effects. An unresolved label is used when the available representation does not support a reliable decision. Only matched and compatible rules proceed to scope inference.

This alignment step is necessary because action names alone are not reliable. The same action may hide different conditions or effects, while different expressions may describe the same transition. Appendix~\ref{app:method-details} evaluates the controlled vocabulary used for this comparison. Unlike nearest-neighbor retrieval~\citep{karpukhin2020dpr,lewis2020rag}, \method{} does not admit a candidate based on similarity alone. The relation label and the inferred scope must both support transfer.

\subsection{M3: Scope Inference}

Rule alignment produces comparable candidate groups. We use the evidence within each group to estimate where a rule remains valid. For candidate rule \(r\), let \(n_r^+\) and \(n_r^-\) denote the numbers of supporting and contradicting observations. We compute the Laplace-smoothed confidence
\begin{equation}
	q_r=
	\frac{n_r^++1}
	{n_r^++n_r^-+2}.
	\label{eq:laplace-q}
\end{equation}
The smoothing term prevents a small evidence set from producing an extreme confidence value.

We also count evidence at the client level. Let \(K_r^+\), \(K_r^-\), and \(K_r^{\mathrm{eligible}}\) denote the supporting, contradicting, and eligible client sets. A client is eligible only when the precondition of rule \(r\) can occur under its configuration. We define
\begin{equation}
	p_r=
	\frac{|K_r^+|}
	{|K_r^{\mathrm{eligible}}|},
	\qquad
	\rho_r^-=
	\frac{|K_r^-|}
	{|K_r^{\mathrm{eligible}}|}.
	\label{eq:scope-stats}
\end{equation}

The same evidence is evaluated within client clusters to identify cluster-specific rules. A rule supported only by its source client remains private. When the evidence does not support any of these assignments, \method{} marks the rule as unresolved. Such rules remain in the federated store for later evidence updates, but they cannot affect execution. Using \(K_r^{\mathrm{eligible}}\) in the denominator also avoids penalizing a rule for clients on which its precondition cannot occur.

\subsection{M4: Gap Detection and Compilation}

Scope inference determines where each federated rule is valid, but a valid rule is used only when the target client lacks local coverage. M4 therefore applies the local-first policy to decide whether federation is needed for the current query. For target client \(k\), let \(\kappa_k(u)\in[0,1]\) denote
the reliability of a local rule, estimated from its observation
support, consistency, or local verification. If \(u\in A_k\)
and \(\kappa_k(u)\geq\tau_{\mathrm{loc}}\), \method{} preserves
the local prediction. Otherwise, the rule is treated as
uncovered or uncertain, and the client consults the
target-compatible candidate set \(F_{\mathrm{FW}}(k,u)\). Otherwise, it searches the target-compatible candidate set
\begin{equation}
	\begin{aligned}
		F_{\mathrm{FW}}(k,u)
		=\bigl\{g\in F:\;&
		\operatorname{rel}(u,g)\in\mathcal{R}_{\mathrm{ok}},\\
		&
		\operatorname{Compat}(k,\sigma(g))=1
		\bigr\},
	\end{aligned}
	\label{eq:fw-runtime-set}
\end{equation}
where
\(\mathcal{R}_{\mathrm{ok}}
=\{\mathrm{match},\mathrm{compatible}\}\).

We define a local gap as a key that is absent from \(A_k\) but has at least one candidate in \(F_{\mathrm{FW}}(k,u)\). When several candidates remain, \method{} ranks them by relation type, scope specificity, and \(q_r\). The compiler rewrites the selected abstract effect into the vocabulary used by target client \(k\). When no candidate passes the filter, the client uses a local backoff if one is available and otherwise abstains. This rule allows compatible knowledge to fill a local gap without exposing the client to an incompatible majority effect. Algorithm~\ref{alg:fedworld_main} summarizes the runtime procedure. 

\begin{algorithm}[t]
	\caption{\method{} prediction and update filtering for target client \(k\)}
	\label{alg:fedworld_main}
	\begin{algorithmic}[1]
		\REQUIRE Local store \(A_k\), federated store \(F\), target client \(k\), query rule \(r\)
		\ENSURE Predicted effect, compiled instruction, or abstention
		\STATE \(u\leftarrow\operatorname{key}(r)\)
		\IF{\(u\in A_k\)}
		\STATE \textbf{return} \(\mathrm{LWM}_k(u)\)
		\ENDIF
		\STATE
		\(\mathcal{C}\leftarrow
		\{g\in F:
		\operatorname{rel}(u,g)
		\in\mathcal{R}_{\mathrm{ok}}\}\)
		\STATE \(\mathcal{C}_k\leftarrow\varnothing\)
		\FORALL{\(g\in\mathcal{C}\)}
		\STATE Infer \(\sigma(g)\) from
		\(K_g^+,K_g^-,K_g^{\mathrm{eligible}}\)
		\IF{\(\operatorname{Compat}(k,\sigma(g))=1\) and \(g\) activates for \(r\)}
		\STATE \(\mathcal{C}_k\leftarrow
		\mathcal{C}_k\cup\{g\}\)
		\ENDIF
		\ENDFOR
		\IF{\(\mathcal{C}_k\neq\varnothing\)}
		\STATE Rank \(\mathcal{C}_k\) by relation, scope specificity, and \(q_g\)
		\STATE \(g^\star\leftarrow\) highest-ranked candidate
		\STATE \textbf{return} \(\operatorname{Compile}_k(g^\star)\)
		\ENDIF
		\IF{\(\mathrm{LWM}_k\) provides a local backoff for \(u\)}
		\STATE \textbf{return} local backoff
		\ENDIF
		\STATE \textbf{return} abstain
	\end{algorithmic}
\end{algorithm}

\section{Theoretical Analysis}
\label{sec:theory}

Let \(F\) be the federated abstract store and let \(u\) be a query from target client \(k\). We define the aligned candidate set as
\begin{equation}
	F_{\mathrm{all}}(k,u)=
	\{g\in F:
	\operatorname{rel}(u,g)
	\in\mathcal{R}_{\mathrm{ok}}\},
	\label{eq:theory-all-set}
\end{equation}
and the scope-compatible subset as
\begin{equation}
	F_{\mathrm{FW}}(k,u)=
	\{g\in F_{\mathrm{all}}(k,u):
	\operatorname{Compat}(k,\sigma(g))=1\}.
	\label{eq:theory-fw-set}
\end{equation}
By construction,
\begin{equation}
	F_{\mathrm{FW}}(k,u)
	\subseteq
	F_{\mathrm{all}}(k,u)
	\label{eq:candidate-containment}
\end{equation}
for every target query.

Let \(\hat y_i^m\in\{0,1\}\) indicate whether method \(m\) predicts evaluation record \(i\) correctly. For a reference method \(b\), paired negative and positive transfer are
\begin{align}
	\mathrm{NTR}(m,b)
	&=
	\frac{1}{n}
	\sum_{i=1}^{n}
	\mathbb{I}
	[\hat y_i^b=1,\hat y_i^m=0],
	\label{eq:theory-ntr}\\
	\mathrm{PTR}(m,b)
	&=
	\frac{1}{n}
	\sum_{i=1}^{n}
	\mathbb{I}
	[\hat y_i^b=0,\hat y_i^m=1].
	\label{eq:theory-ptr}
\end{align}

We further define
\begin{equation}
	\mathrm{NetTransfer}(m,b)
	=
	\mathrm{PTR}(m,b)
	-
	\mathrm{NTR}(m,b).
	\label{eq:theory-net-transfer}
\end{equation}
Let \(p_{+}\) denote the fraction of cases where \(b\) is wrong and a correct compatible rule is available, and \(p_{-}\) the fraction where \(b\) is correct but an incompatible rule may overwrite it. For method \(m\), \(\beta_m\) is the rate of withholding correct compatible rules, while \(\alpha_m\) is the rate of activating incompatible rules.

\begin{theorem}[Protection--transfer trade-off]
	\label{thm:transfer-tradeoff}
	Fix the abstraction map, relation classifier, candidate
	ordering, and compiler. Assume that admitting a correct
	scope-compatible rule resolves a transferable record, while
	an error on a protected record occurs exactly when a
	scope-incompatible rule is activated. Then
	\begin{align}
		\mathrm{PTR}(m,b)
		&=
		(1-\beta_m)p_{+},
		\label{eq:ptr-tradeoff}\\
		\mathrm{NTR}(m,b)
		&=
		\alpha_m p_{-},
		\label{eq:ntr-tradeoff}\\
		\mathrm{NetTransfer}(m,b)
		&=
		(1-\beta_m)p_{+}
		-
		\alpha_m p_{-}.
		\label{eq:net-tradeoff}
	\end{align}
	Consequently, \method{} achieves higher net transfer than
	NaiveUnion when
	\begin{equation}
		(\alpha_{\mathrm{NU}}-\alpha_{\mathrm{FW}})p_{-}
		>
		(\beta_{\mathrm{FW}}-\beta_{\mathrm{NU}})p_{+}.
		\label{eq:tradeoff-condition}
	\end{equation}
\end{theorem}

Theorem~\ref{thm:transfer-tradeoff} makes the cost of scope
filtering explicit. A stricter filter is beneficial when the
harmful overwrites it prevents exceed the useful transfers it
withholds. Thus, the objective is not to minimize negative
transfer alone, since a method could do so by rejecting all
federated rules. Instead, \method{} should reduce incompatible
activations while retaining enough compatible rules to obtain
positive net transfer. The proof is provided in
Appendix.

\section{Experiments}
\label{sec:experiments}

\paragraph{Benchmarks and client construction.}
We evaluate \method{} on $\tau$-bench and ALFWorld. $\tau$-bench contains retail and airline interactions governed by different business policies~\citep{yao2025taubench}. As a result, the same tool action may produce different outcomes across policy profiles. ALFWorld provides text-based household tasks with physical and symbolic state transitions~\citep{shridhar2021alfworld}. Its action effects may vary with object states, receptacle constraints, and environment layouts. The two benchmarks therefore cover policy-dependent and environment-dependent transition conflicts.
We construct 11 $\tau$-bench clients from six retail and five airline policy profiles, and 12 ALFWorld clients from distinct household transition profiles. 

\paragraph{Data partition and evaluation splits.}
We partition abstract identity keys rather than raw trajectories. For each client, 60\% of eligible keys are used for the local rule store, 20\% for validation, and 20\% for held-out evaluation, preventing the same normalized rule from appearing in both local and cross-client test sets.
The offline evaluation includes six splits. S0 covers locally available rules, S3 tests cross-client gap filling, and S4 contains conflict cases. $S4_{strict}$ denotes cases where the target effect differs from the federated majority, while $S4_{trivial}$ denotes agreement with the majority. S5 mixes held-out local and cross-client cases. Full split definitions and record counts are provided in Appendix. We select
$
|K_r^+|\geq 3,
p_r\geq 0.60,
\rho_r^-\leq 0.10
$
on the validation set and use the same values for both benchmarks. New domains may recalibrate them on a small validation set. Results are averaged over five fixed client-partition seeds.

\paragraph{Comparison methods.}
We evaluate six method configurations, denoted B0--B5. B0 uses no world model and measures the fixed controller without stored transition knowledge. B1, denoted \localwm{}, uses only the target client's detailed local experience. B2 keeps the target client's normalized abstract rules but introduces no cross-client sharing. B3, denoted \naiveunion{}, pools abstract rules from all clients without relation alignment or scope filtering \cite{xu2025amem}. B4 aligns related cross-client rules but does not check whether their scope is compatible with the target client \cite{chen2026fedagent}. B5 is the complete \method{} protocol, including abstract rule mapping, cross-client alignment, scope inference, and local-first compilation.
We also report centralized detailed retrieval (CDR). CDR searches pooled detailed records from all clients and therefore has access to information that remains local under the federated methods. We use it as a centralized reference rather than a privacy-preserving baseline. 

\paragraph{Offline evaluation.}
Offline evaluation measures abstract next-state prediction using exact match after deterministic normalization. We also report paired positive and negative transfer to distinguish useful gap filling from harmful overwrite. Given method \(m\) and reference method \(b\),
$
	\mathrm{PTR}(m,b)
	=
	\frac{1}{N}
	\sum_{i=1}^{N}
	\mathbb{I}[b_i=0\land m_i=1],
	\notag$$
	\mathrm{NTR}(m,b)
	=
	\frac{1}{N}
	\sum_{i=1}^{N}
	\mathbb{I}[b_i=1\land m_i=0].
	\label{eq:transfer-rates}
$
 \(\mathrm{PTR}\) counts reference errors corrected by method \(m\), whereas \(\mathrm{NTR}\) counts correct reference predictions broken by method \(m\). We further report
\(\mathrm{NetTransfer}=\mathrm{PTR}-\mathrm{NTR}\).
Unless otherwise stated, B1 serves as the reference because it represents the target client's original local knowledge.

\paragraph{Online evaluation.}
We execute all online episodes with the same rule-conditioned
controller and action budget. State regression counts actions whose
observed post-state reverses the predicted transition. Repeated action
counts consecutive executions of the same action argument pair
without an intervening state change. Excess steps measures the number
of actions beyond the shortest valid plan, while task success is
determined by the benchmark terminal predicate. We execute 400
$\tau$-bench episodes and 600 ALFWorld episodes for each partition
seed.
\paragraph{Additional results.}
The appendix reports the complete split level results, client and record statistics, seed level stability, scope distribution, alignment analysis, and representative conflict cases. It also provides the detailed experimental protocol and supporting proofs.
\begin{table}[t]
	\centering
	\small
	\begin{tabular*}{\columnwidth}
		{@{\extracolsep{\fill}}lrrrrr@{}}
		\toprule
		Method
		& S0
		& S3
		& S4$_{\mathrm{strict}}$
		& S4$_{\mathrm{trivial}}$
		& S5 \\
		\midrule
		
		\multicolumn{6}{@{}l}{\textit{$\tau$-bench}} \\
		B0
		& 0.332 & 0.254 & 0.306 & 0.221 & 0.246 \\
		B1
		& \textbf{0.846} & 0.712 & 0.694 & 0.731 & 0.681 \\
		B2
		& 0.804 & 0.665 & 0.625 & 0.702 & 0.640 \\
		B3
		& 0.821 & 0.794 & 0.556 & 0.913 & 0.752 \\
		B4
		& 0.836 & 0.812 & 0.611 & \textbf{0.932} & 0.773 \\
		B5
		& \textbf{0.846} & \textbf{0.836} & \textbf{0.722}
		& 0.924 & \textbf{0.803} \\
		CDR
		& 0.828 & 0.807 & 0.583 & 0.918 & 0.781 \\
		
		\midrule
		\multicolumn{6}{@{}l}{\textit{ALFWorld}} \\
		B0
		& 0.284 & 0.211 & 0.243 & 0.228 & 0.205 \\
		B1
		& \textbf{0.812} & 0.574 & 0.653 & 0.704 & 0.548 \\
		B2
		& 0.759 & 0.528 & 0.583 & 0.661 & 0.497 \\
		B3
		& 0.774 & 0.724 & 0.431 & 0.906 & 0.688 \\
		B4
		& 0.796 & 0.752 & 0.493 & \textbf{0.928} & 0.716 \\
		B5
		& \textbf{0.812} & \textbf{0.786} & \textbf{0.681}
		& 0.922 & \textbf{0.748} \\
		CDR
		& 0.789 & 0.744 & 0.458 & 0.914 & 0.721 \\
		
		\bottomrule
	\end{tabular*}
	\caption{Main offline exact-match results over five client-partition
		seeds. Best results are shown in bold.}
	\label{tab:main_offline}
\end{table}
\subsection{Offline World-Model Accuracy}
\label{sec:offline_results}

Table~\ref{tab:main_offline} reports the main offline results.
B1 performs well on S0 because these rules are already present in the
target client's local store.
Its accuracy falls from 0.846 to 0.712 on $\tau$-bench and from 0.812
to 0.574 on ALFWorld when evaluation moves from S0 to S3.
The drops of 13.4 and 23.8 percentage points confirm that local
experience is useful but incomplete.
B5 preserves the B1 result on S0 because its local-first policy does
not replace an existing local rule.
The benefit appears when local coverage is absent.
On S3, B5 improves over B1 by 12.4 points on $\tau$-bench and
21.2 points on ALFWorld.
The gains remain on the held-out S5 split, where B5 improves by
12.2 and 20.0 points.
These results show that scope filtering does not prevent useful
cross-client transfer. The results on S4 reveal why unrestricted sharing is unsafe.
B3 performs well when the target agrees with the majority, but its
accuracy falls below B1 on strict conflicts.
Relation alignment in B4 reduces some errors, yet B4 remains below B1
on $S4_{strict}$.
B5 obtains the highest strict-conflict accuracy on both benchmarks.
On $S4_{trivial}$, B4 is only 0.8 and 0.6 points above B5.
This small difference is the cost of withholding uncertain rules, while
the strict-conflict gains are substantially larger.
CDR is not consistently better than the federated methods.
Although it accesses detailed pooled records, it does not prevent
client-specific transitions from being retrieved for the wrong target.
The comparison indicates that access to more detailed experience does
not replace target-conditioned scope control.

\begin{table}[t]
	\centering
	\small
	\setlength{\tabcolsep}{1mm}
	\begin{tabular}{@{}lrrrrrrr@{}}
		\toprule
		Benchmark & $n$ & B1 & B3 & B5
		& PTR & NTR & Net \\
		\midrule
		$\tau$-bench
		& 72
		& 0.694
		& 0.556
		& \textbf{0.722}
		& 0.056
		& 0.194
		& $-0.138$ \\
		
		ALFWorld
		& 144
		& 0.653
		& 0.431
		& \textbf{0.681}
		& 0.028
		& 0.250
		& $-0.222$ \\
		\bottomrule
	\end{tabular}
	\caption{Strict-conflict transfer relative to B1. B1, B3, and B5
		report exact match. }
	\label{tab:strict_transfer}
\end{table}

\begin{figure}[!t]
	\centering
	\includegraphics[width=.98\linewidth]
	{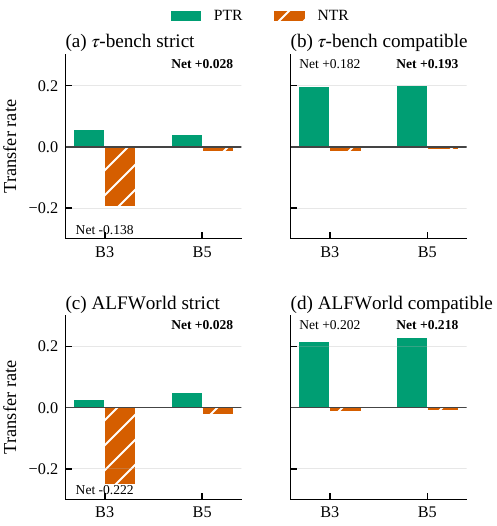}
	\caption{Paired transfer under conflicting dynamics.}
	\label{fig:ntr_transfer_diagnostic}
\end{figure}
\subsection{Scope Filtering under Conflicting Dynamics}
\label{sec:conflict_results}

The strict-conflict split isolates cases in which the federated
majority disagrees with the correct target effect.
Table~\ref{tab:strict_transfer} shows that B3 loses 13.8 EM points
relative to B1 on $\tau$-bench and 22.2 points on ALFWorld.
Its negative-transfer rate is much higher than its positive-transfer
rate, giving net transfer values of \(-0.138\) and \(-0.222\).
B5 recovers 16.6 points over B3 on $\tau$-bench and 25.0 points on
ALFWorld.
It also exceeds B1 by 2.8 points on both benchmarks.
The gain over B1 comes from cluster-specific and private rules that
safely fill cases not resolved by the local lookup.
Thus, local-first prediction does not reduce B5 to a local-only model.

Figure~\ref{fig:ntr_transfer_diagnostic} separates positive and
negative transfer for strict-conflict and compatible-majority cases.
B3 transfers useful rules when the target follows the majority, but
the same policy produces substantial negative transfer for minority
clients.
B5 retains the compatible gain while sharply reducing the overwrite
risk.
Scope filtering therefore changes which rules may be used rather than
disabling federation.

\subsection{Online Execution Quality}
\label{sec:online_results}

\begin{figure}[!t]
	\centering
	\includegraphics[width=.98\linewidth]
	{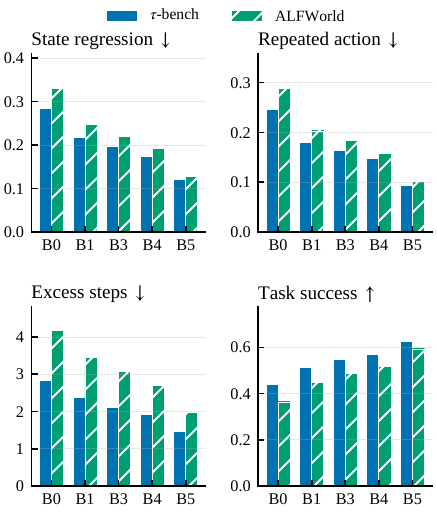}
	\caption{Online execution results.}
	\label{fig:h3online}
\end{figure}

A correct abstract transition is useful only when it improves the
agent's actions.
Table~\ref{tab:micro_all} and Figure~\ref{fig:h3online} report the
online results under the fixed controller.
Relative to B1, B5 reduces state regression by 44.5\% on
$\tau$-bench and 48.2\% on ALFWorld.
Repeated actions fall by 48.1\% and 50.2\%, while excess steps fall by
38.4\% and 42.8\%.
Task success increases by 11.2 and 14.6 percentage points.
B3 and B4 improve over B1 on average because they provide additional
cross-client rules.
However, their remaining conflict errors lead to more regressions,
repeated actions, and excess steps than B5.
The online results connect scope errors to agent behavior: an
incompatible transition does not cause only an offline prediction
mistake, but can also send the controller into repeated or regressive
actions.

\begin{table}[!t]
	\centering
	\small
	\begin{tabular*}{\columnwidth}{@{\extracolsep{\fill}}lrrrrr@{}}
		\toprule
		Metric & B0 & B1 & B3 & B4 & B5 \\
		\midrule
		
		\multicolumn{6}{@{}l}{\textit{$\tau$-bench}} \\
		State regression $\downarrow$
		& 0.284 & 0.218 & 0.196 & 0.174 & \textbf{0.121} \\
		Repeated action $\downarrow$
		& 0.246 & 0.181 & 0.164 & 0.147 & \textbf{0.094} \\
		Excess steps $\downarrow$
		& 2.84 & 2.37 & 2.12 & 1.93 & \textbf{1.46} \\
		Task success $\uparrow$
		& 0.438 & 0.512 & 0.548 & 0.571 & \textbf{0.624} \\
		
		\midrule
		\multicolumn{6}{@{}l}{\textit{ALFWorld}} \\
		State regression $\downarrow$
		& 0.331 & 0.247 & 0.219 & 0.192 & \textbf{0.128} \\
		Repeated action $\downarrow$
		& 0.289 & 0.203 & 0.184 & 0.158 & \textbf{0.101} \\
		Excess steps $\downarrow$
		& 4.18 & 3.46 & 3.08 & 2.71 & \textbf{1.98} \\
		Task success $\uparrow$
		& 0.364 & 0.447 & 0.486 & 0.518 & \textbf{0.593} \\
		
		\bottomrule
	\end{tabular*}
	\caption{Online execution under the fixed rule-conditioned
		controller.}
	\label{tab:micro_all}
\end{table}

\subsection{Stability across Client Partitions}

\begin{figure}[!t]
	\centering
	\includegraphics[width=.98\linewidth]
	{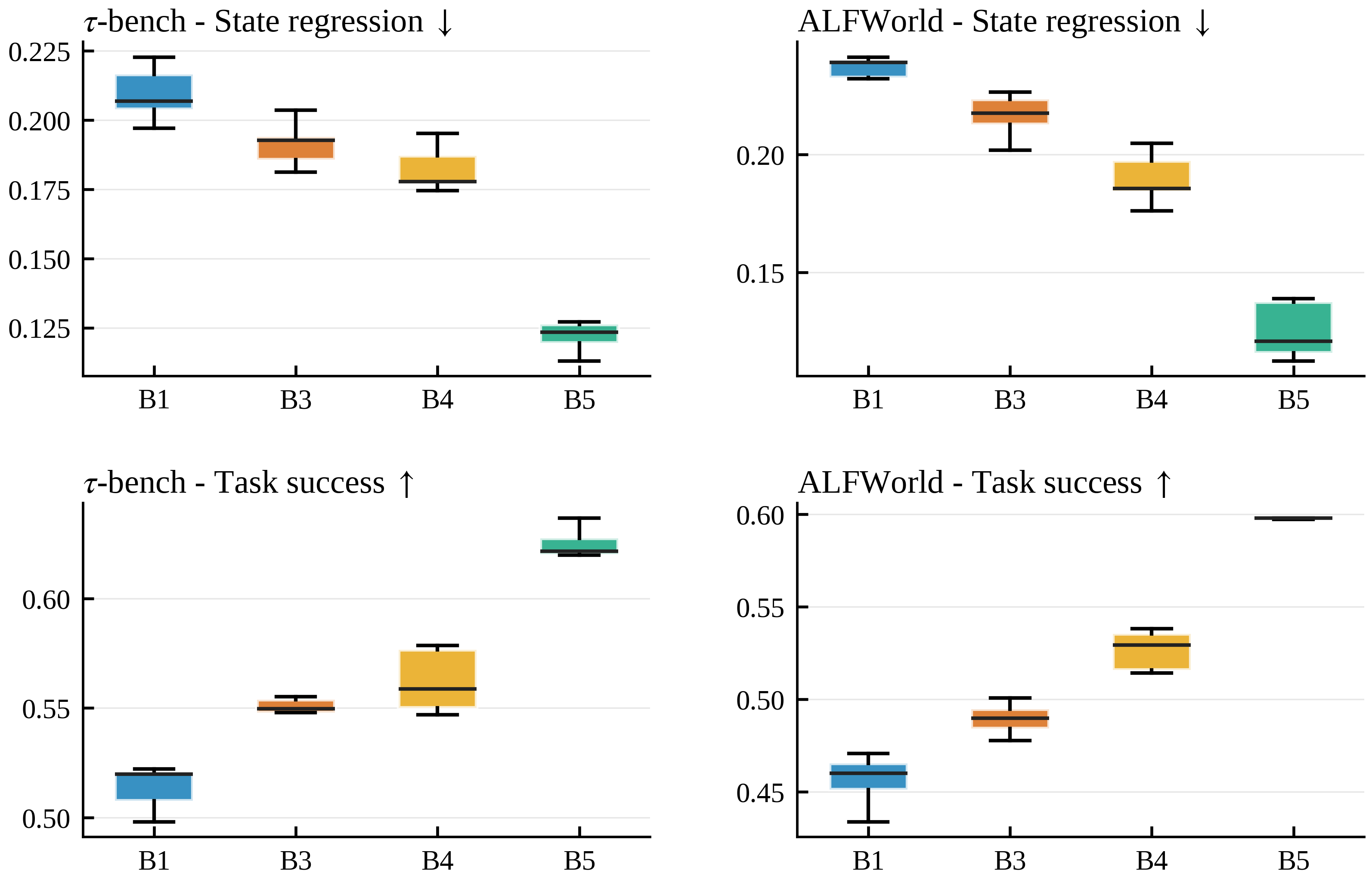}
	\caption{Per-seed execution stability.}
	\label{fig:micro_metrics_diagnostic}
\end{figure}

Figure~\ref{fig:micro_metrics_diagnostic} shows the corresponding
per-seed distributions for representative online metrics.
The advantage of B5 appears across the partition seeds rather than
being produced by one favorable split.
On both benchmarks, its median state regression rate is lower, while its median task success is higher.
The spread of B5 remains comparable to or smaller than the competing
methods.
Scope filtering therefore does not obtain its average improvement by
accepting unstable rules on a small number of client assignments.
The consistent direction across seeds is especially important for
ALFWorld, where client profiles contain wider variation in object
states and transition preconditions.

\section{Conclusion}
\label{sec:limits}

We introduced \method{}, a scope-aware federated world model that preserves local rules while enabling safe cross-client transfer. A limitation is that \method{} assumes sufficiently specified, deterministic abstract transitions and does not yet model partial observability, stochastic effects, or temporally extended actions. Future work will learn this abstraction automatically for web and software agents while retaining reliable abstention under uncertain evidence.
	
	\bibliography{references}
	
	\clearpage
	\appendix

\end{document}